\documentclass[11pt,a4paper]{article}
\pdfoutput=1
\usepackage{jinstpub}

\usepackage{lineno}

\begin{document}
\hyphenation{had-ron-i-za-tion}
\hyphenation{cal-or-i-me-ter}
\hyphenation{de-vices}
\title{
  Performance of a Triple-GEM Demonstrator in $pp$ Collisions at the CMS Detector
}

\author[n]{M.~Abbas,} 
\author[s]{M.~Abbrescia,}
\author[h,j]{H.~Abdalla,}
\author[h,k]{A.~Abdelalim,}
\author[h,i]{S.~AbuZeid,}
\author[d]{A.~Agapitos,}
\author[ad]{A.~Ahmad,}
\author[q]{A.~Ahmed,}
\author[ad]{W.~Ahmed,}
\author[w]{C.~Aim\`e,}

% ...
\affiliation[d]{Peking University, Beijing, China} %
% ...
\affiliation[h]{Academy of Scientific Research and Technology - ENHEP, Cairo, Egypt} %
\affiliation[i]{Ain Shams University, Cairo, Egypt}
\affiliation[j]{Cairo University, Cairo, Egypt}
\affiliation[k]{Helwan University, also at Zewail City of Science and Technology, Cairo, Egypt}
% ...
\affiliation[n]{Karlsruhe Institute of Technology, Karlsruhe, Germany}
% ...
\affiliation[q]{Delhi University, Delhi, India}
% ...
\affiliation[s]{Politecnico di Bari, Universit\`{a} di Bari and INFN Sezione di Bari, Bari, Italy}%
% ...
\affiliation[w]{Universit\`{a} di Pavia and INFN Sezione di Pavia, Pavia, Italy} %
% ...
\affiliation[ad]{National Center for Physics, Islamabad, Pakistan}
% ...

\emailAdd{ian.james.watson@cern.ch}

\date{\today}

\abstract{
After the Phase-2 high-luminosity upgrade to the Large Hadron Collider (LHC), the collision rate and therefore the background rate will significantly increase, particularly in the high $\eta$ region.
To improve both the tracking and triggering of muons, the Compact Muon Solenoid (CMS) Collaboration plans to install triple-layer Gas Electron Multiplier (GEM) detectors in the CMS muon endcaps.
Demonstrator GEM detectors were installed in CMS during 2017 to gain operational experience and perform a preliminary investigation of detector performance.
We present the results of triple-GEM detector performance studies performed in situ during normal CMS and LHC operations in 2018.
The distribution of cluster size and the efficiency to reconstruct high $p_T$ muons in proton--proton collisions  are presented as well as the measurement of the environmental background
rate to produce hits in the GEM detector.
}

\keywords{Muon Spectrometers, Micropattern gaseous detectors, Performance of High Energy Physics Detectors}

\hypersetup{%
pdfauthor={Dajeong Jeon, Teruki Kamon, Yechan Kang, Jason Lee, Ian J. Watson [ed.]},%
pdftitle={Performance of a Demonstrator Triple-GEM Detector in pp Collisions at the CMS Detector},%
pdfsubject={CMS},%
pdfkeywords={CMS, physics, detector, GEM, performance}}

\maketitle 

\section{Introduction}

The Compact Muon Solenoid (CMS) detector \cite{Chatrchyan:2008zzk} at the Large Hadron Collider (LHC) has performed remarkably during the running of the LHC. 
CMS will be upgraded for the High Luminosity phase of the LHC (HL-LHC)~\cite{Contardo:2020886}, which will deliver an order of magnitude increase in the instantaneous luminosity to \(5 \times 10^{34}\) cm\(^{-2}\) s\(^{-1}\) and the center-of-mass energy may increase from 13~TeV to 14~TeV.
The corresponding increase in the collision rate, and therefore
radiation background, will cause difficulties in maintaining the high efficiency and reliability of the data collection triggers with the current CMS detector configuration.
In particular, final states with muons are extremely important to the CMS physics program, providing the motivation to the CMS Collaboration to upgrade the muon systems to cope with the high-luminosity conditions.
The high-\(\eta\) region is of particular concern, as the increased luminosity will deliver additional radiation primarily in this forward region.
Therefore, to enhance the muon trigger and track reconstruction capabilities, large-area triple-layer Gas Electron Multiplier (GEM) detectors \cite{SAULI1997531} are being installed in the CMS muon endcaps \cite{Colaleo:2021453}.

\begin{figure}
  \centering
  \includegraphics[height=.22\textheight]{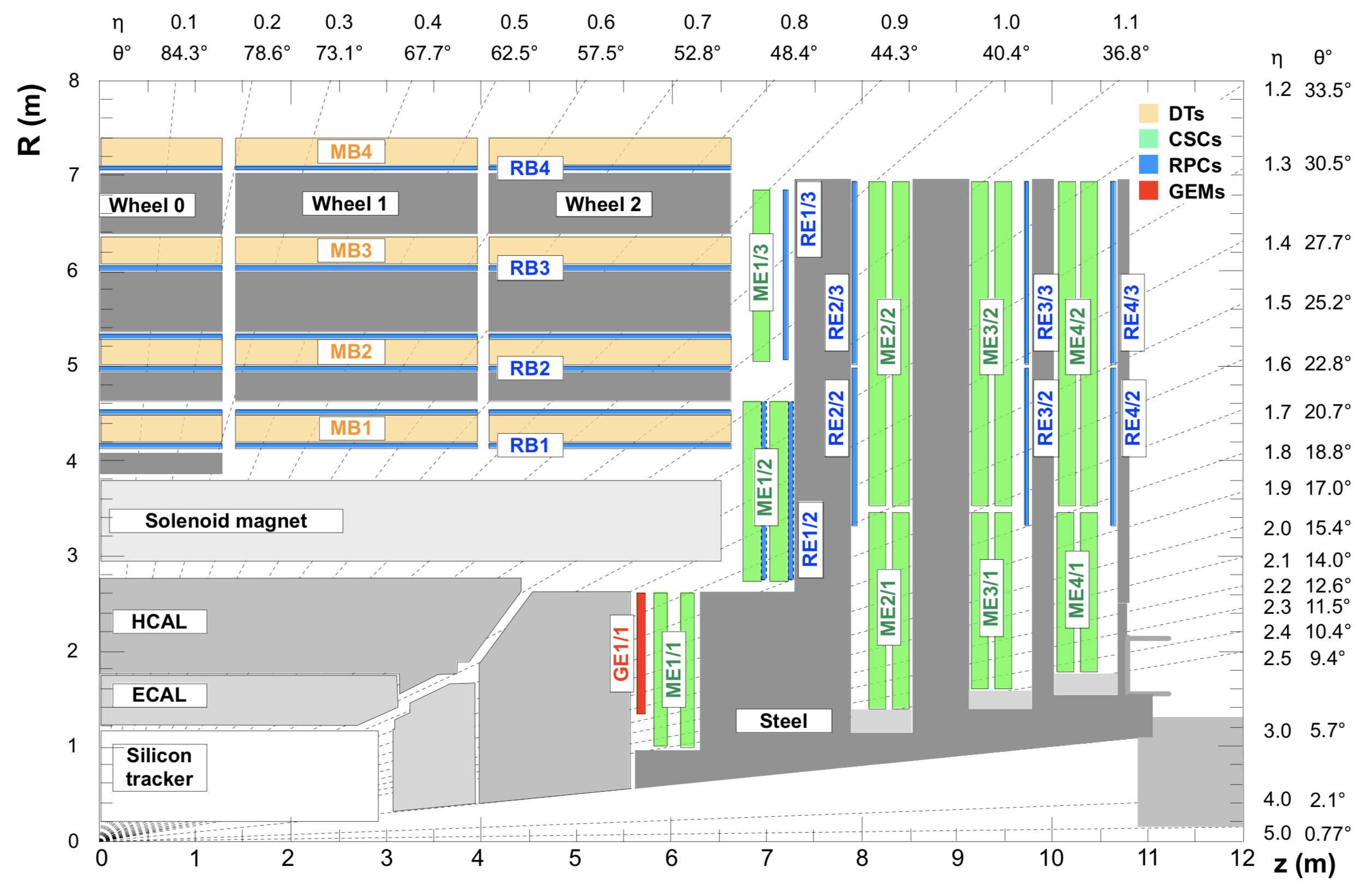}
  \includegraphics[height=.22\textheight]{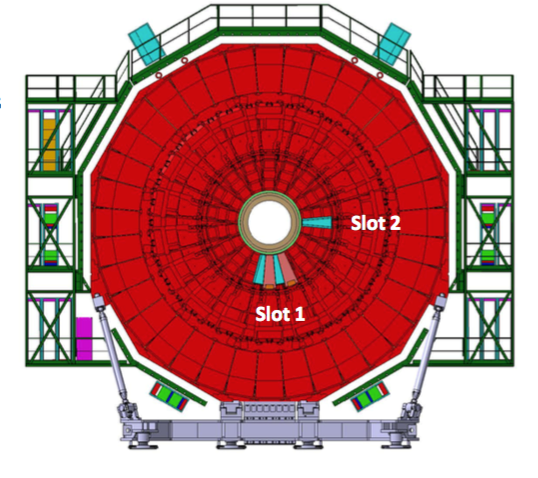}
  \caption{Figure showing a quadrant of the $R$-$z$ cross-section of the CMS detector with the position of the GE1/1 detector highlighted in red (left), and an $x$-$y$ view showing the locations where the demonstrator detectors were installed, showing the 4 chambers used in the studies in slot 1, and the chamber with upgraded electronics that was not included in the data taking in slot 2 (right).}
  \label{fig:ge11-rz}
\end{figure}

The GEM detectors are designed to complement the existing Cathode Strip Chambers (CSC).
One station of GEM detectors was installed in 2019--2020  during the LHC Long Shutdown 2.
The most forward station of CSC detectors is denoted ME1/1, and the GEM detectors were installed in front of ME1/1 and are called GE1/1, as shown in figure~\ref{fig:ge11-rz}.
Two more stations of GEM detectors will be installed during the next LHC long shutdown in preparation of the HL-LHC. Further information on the design and geometry of the GEM detectors is provided in the Technical Design Report~\cite{Colaleo:2021453}.

The GE1/1 station at each muon endcap consists of 36 super chambers, each covering a \(\phi\) slice of $10.15^\circ$ and arranged in an overlapping fashion to provide full $2\pi$ coverage in $\phi$.
The super chambers alternate in $\phi$ between a long version covering $1.55 < |\eta| < 2.18$ and a short version covering $1.61 < |\eta| < 2.18$, as required by the mechanical envelope of the existing endcap, while providing maximal $\eta$ coverage.
A super chamber is comprised of two layers of trapezoidal triple-GEM detectors, each of which provides an independent measurement of particle position.
Each layer consists of eight $\eta$ partitions of 384 strips per partition.
The strips are arranged to give good resolution in the global $\phi$ direction, and the partitioning gives a coarse measurement of the $\eta$ of a particle traversing the detector.
The 384 strips are read out by three Very Forward ATLAS and TOTEM (VFAT) front-end readout chips, each responsible for outputting a binary value from 128 channels indicating if the corresponding strip was hit.

Preparation for the utilization of GE1/1 began with the insertion of \emph{demonstrator} super chambers during the 2016 year end technical shutdown (YETS) \cite{Abbas_2020}.
Four superchambers were installed vertically in slot 1 as shown in figure~\ref{fig:ge11-rz}, and one superchamber was installed horizontally in slot 2.
The chambers used different versions of the detector readout electronics; the detectors in slot 1 use version 2 which incorporate VFAT2 ASICs~\cite{vfat2}, while the detector in slot 2 was equipped with version 3 which uses VFAT3 ASICs~\cite{vfat3}.
The integration with the CMS DAQ system was only available for the version 2 detectors.
The electronics chains, DAQ, High Voltage system and detector control system used during the slice test are further described in ref.~\cite{Abbas_2020}.
The main goal of this GEM demonstrator detector was to gain operational experience with running the GEM detector and to understand the performance of the detector during proton-proton collisions.

The detector was operated at a gas gain of \(\sim 1 \times 10^4\) with a readout charge threshold of 3 fC for eta partitions three to eight.
With this setting, a measured detection efficiency of 95\% for muons is consistent with the test beam data~\cite{ABBAS2020164104}.  
Eta partitions one and two were operated at higher thresholds, 8.8 fC and 4.0 fC respectively, due to noise from a grounding issue in the readout electronics of version 2.
This noise problem is fixed in version 3 of the electronics.
The discharge rate and gain variation due to a potential drop caused by the current load are evaluated in the TDR~\cite{Colaleo:2021453}.
We did not expect such effects in the demonstrator where chambers were operated at a lower gain of \(1 \times 10^4\).
However, irreversible channel loss due to discharge was observed in the demonstrator chambers.
The subsequent investigation provided a strategy for mitigating the effect of these discharge events: increasing the protection resistor of each VFAT channel and the decoupling resistor in the chamber HV filter~\cite{Ivone:2020jnk}.

In this paper, we present the results of the GEM demonstrator detector's performance using data collected during normal CMS operations with LHC proton--proton collisions.
This includes validation of the GEM reconstruction software chain with real data, environmental background rate studies, and muon detection efficiency studies.
Although the GEM demonstrator detectors were installed and operational from the end of 2017, they were still being commissioned.
To avoid disrupting the CMS read-out system during stable collisions, GEM was only included in a subset of luminosity runs. 
We present results from data collected on July 8th, 2018, when the GEM detector was incorporated into the full CMS DAQ chain, corresponding to an integrated luminosity of 205.4~pb$^{-1}$.

\begin{figure}[t]
  \centering
  \includegraphics[width=.9\textwidth]{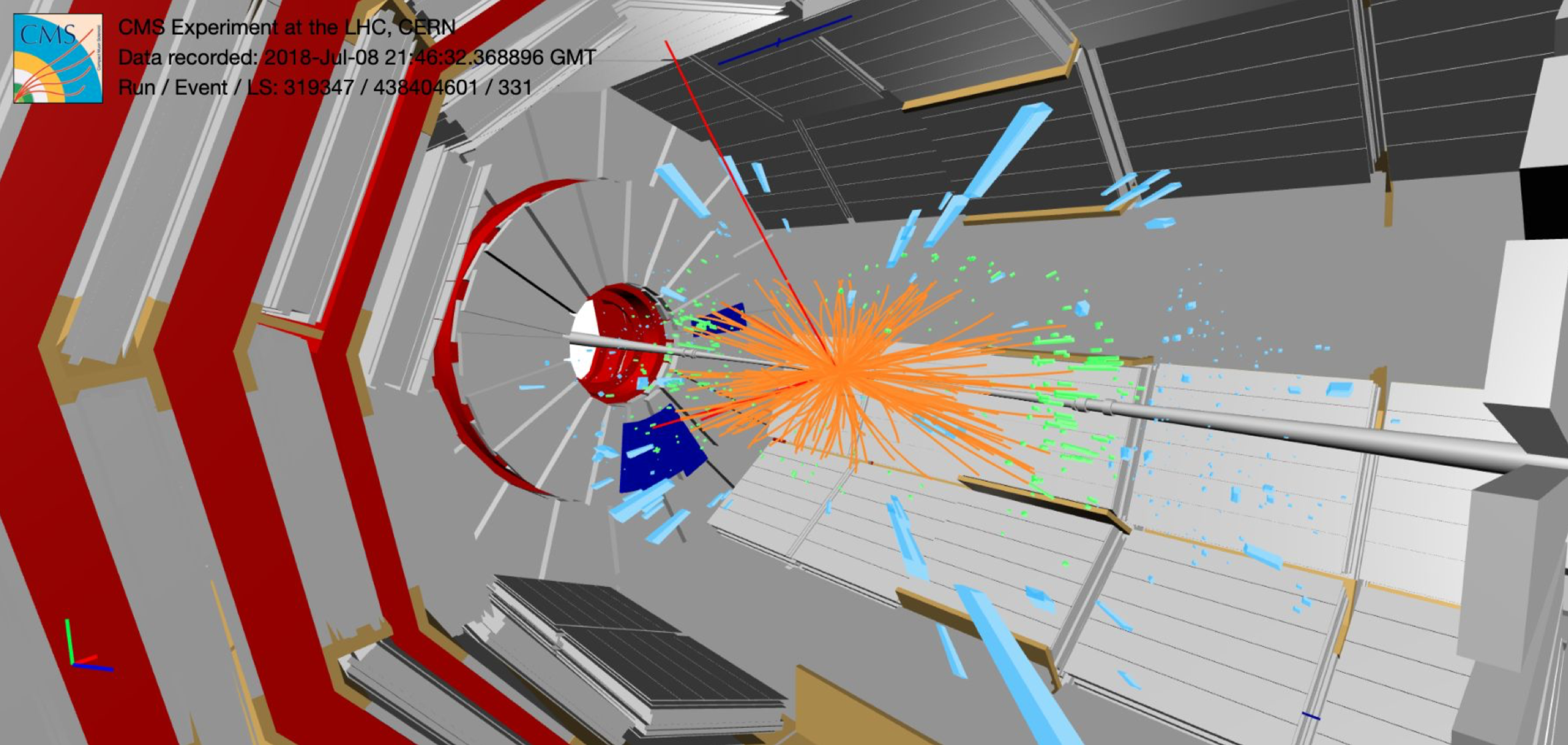}

  \includegraphics[width=.45\textwidth]{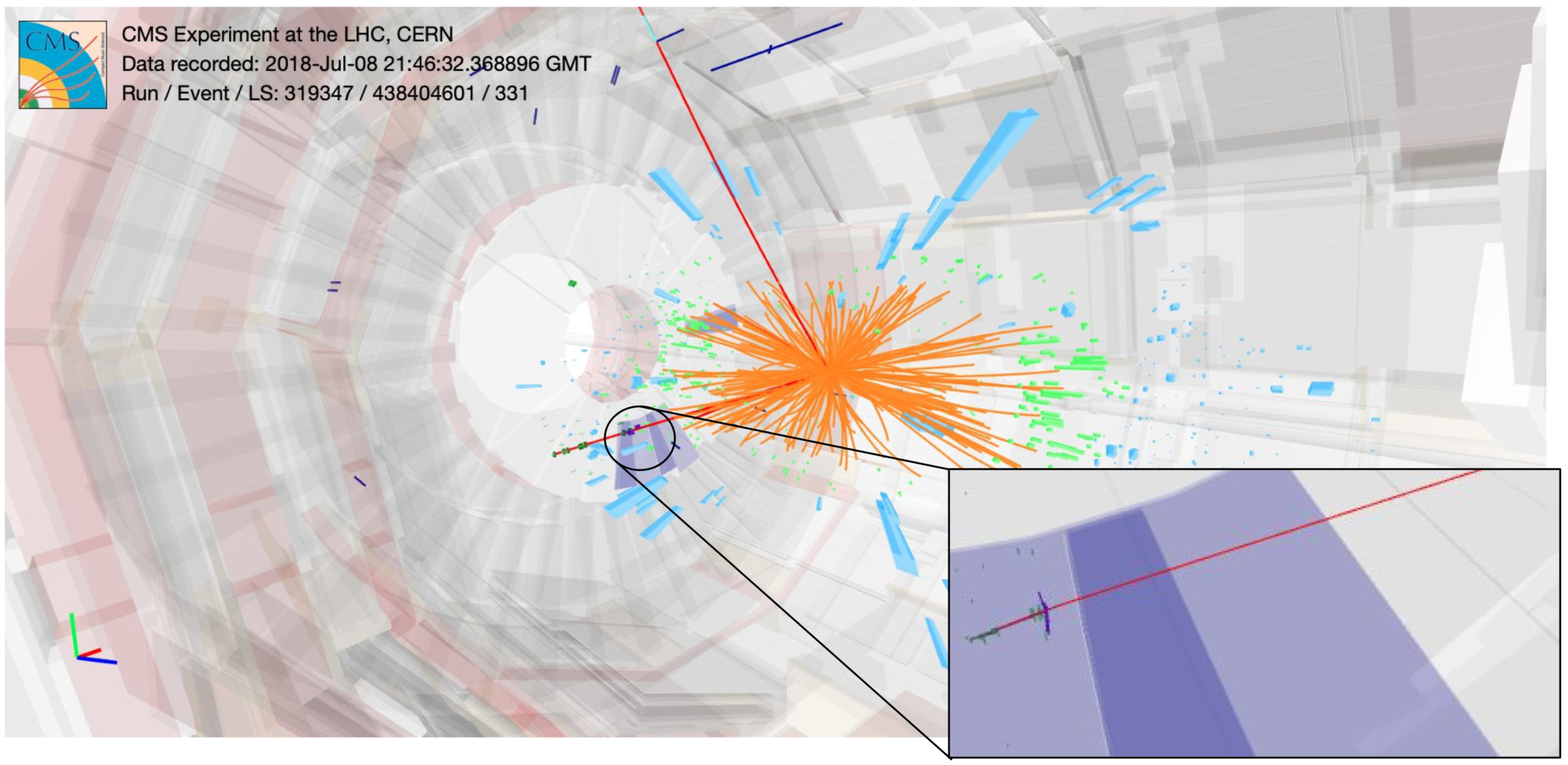}
  \includegraphics[width=.45\textwidth]{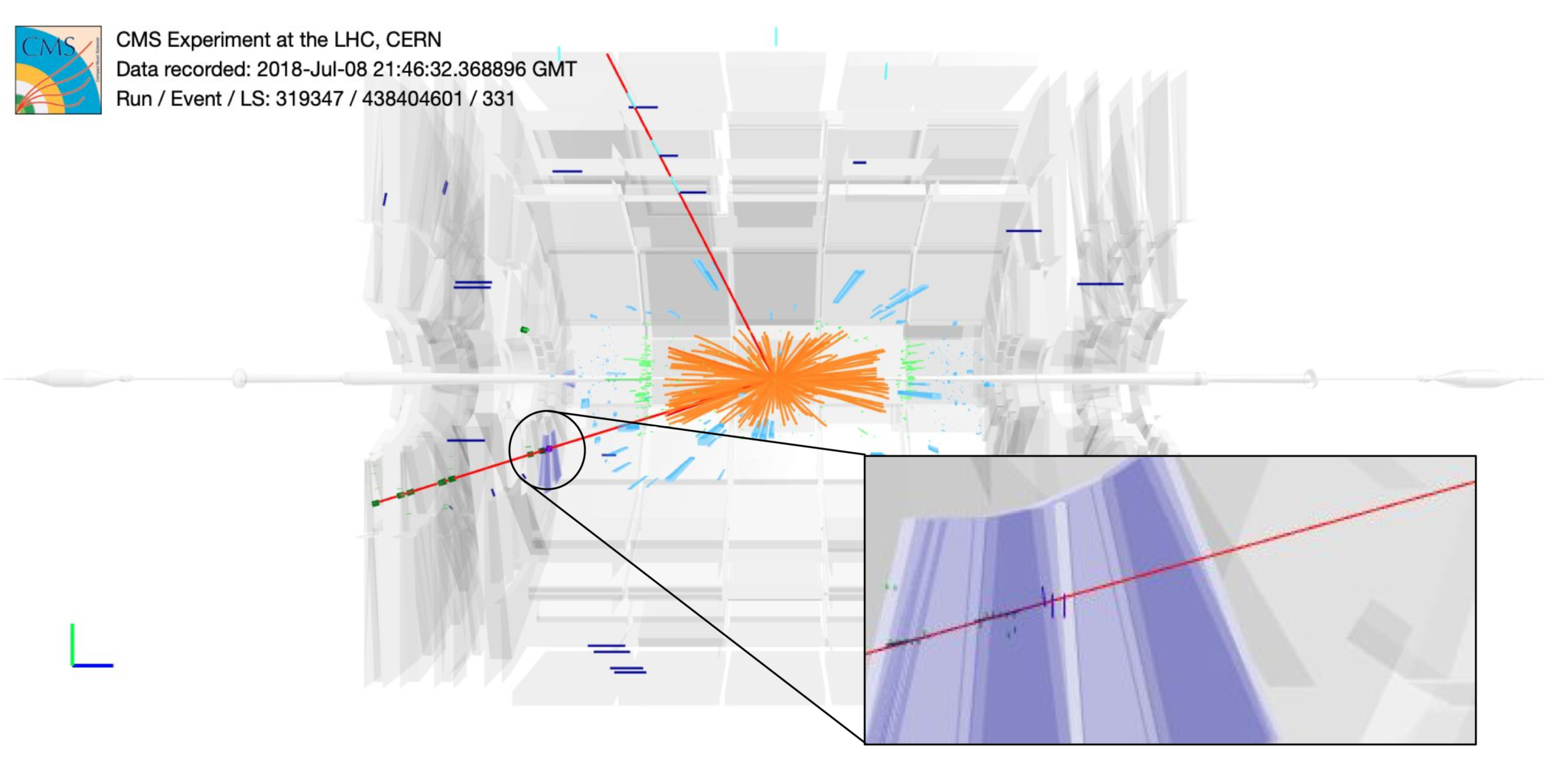}
  \caption{Event display showing a reconstructed muon (the red lines) passing through the GEM detector (the blue trapezoidal boxes) where a GEM hit has been reconstructed along the muon path in $R$-$z$ view (bottom left) and azimuthal view (bottom right).}
  \label{fig:event-display}
\end{figure}

In addition to providing a striking visual representation of particle physics collision data, event displays can be a useful debugging tool, particularly for early running.
Therefore, we have updated the iSpy program~\cite{alverson2012ispy} used to display CMS events to include the GEM demonstrator detectors.
Figure~\ref{fig:event-display} shows an example event display where a GEM hit has been found along the path of a well-reconstructed muon belonging to a $Z$ boson candidate which decays to two muons.

\section{Distribution of Cluster Size}

\begin{figure}[t]
  \centering
  \includegraphics[width=.24\textwidth]{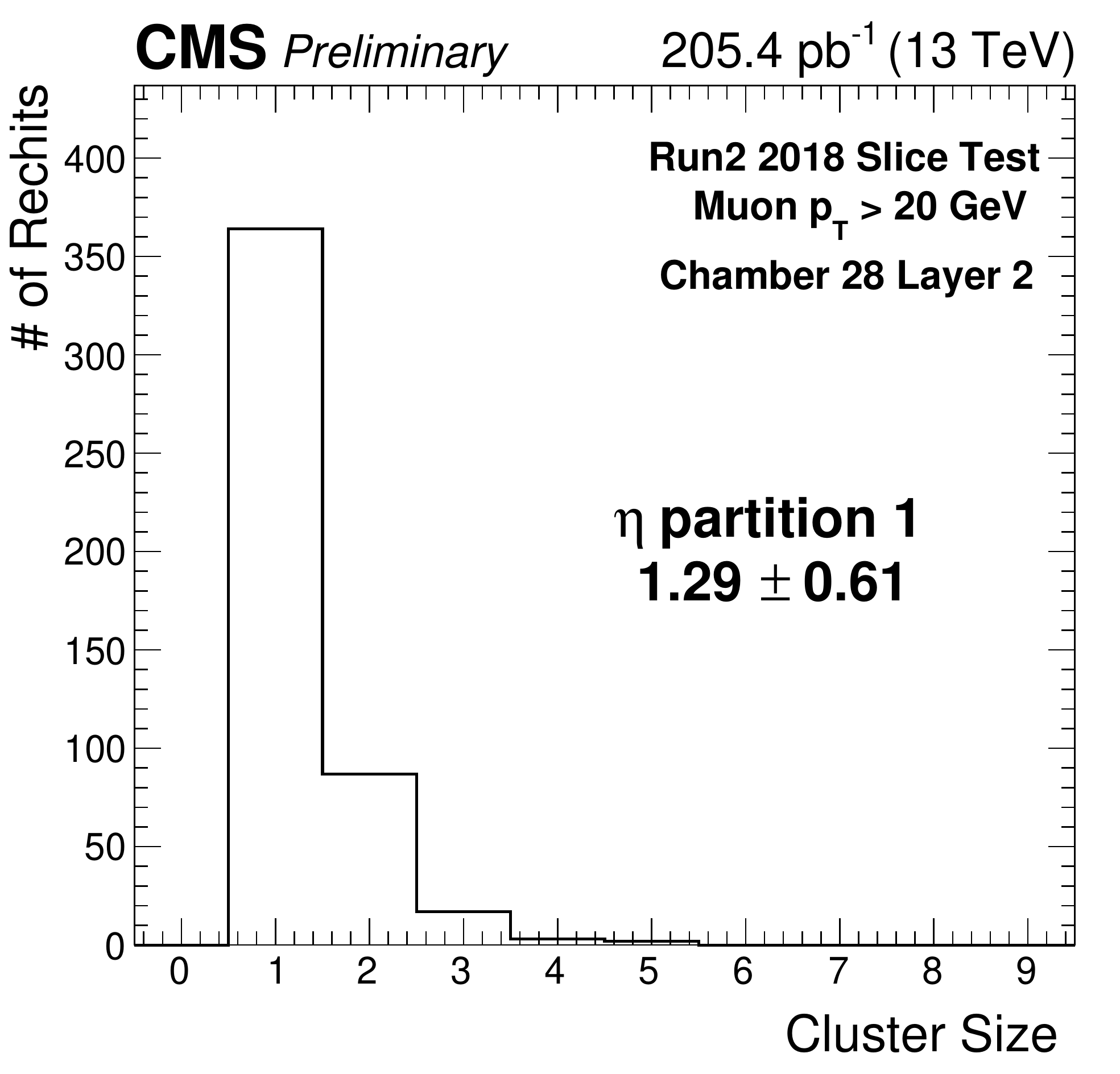}
  \includegraphics[width=.24\textwidth]{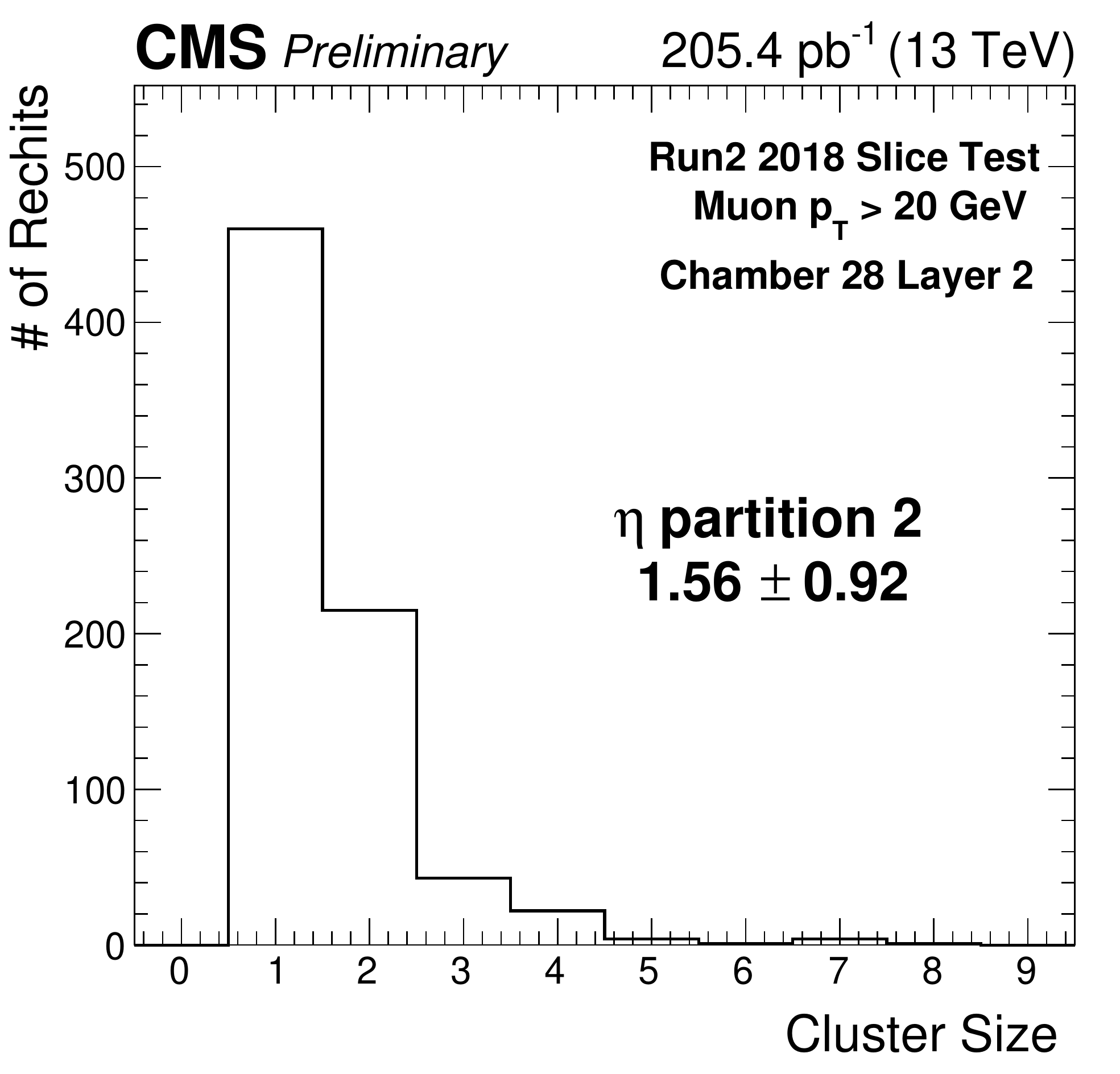}
  \includegraphics[width=.24\textwidth]{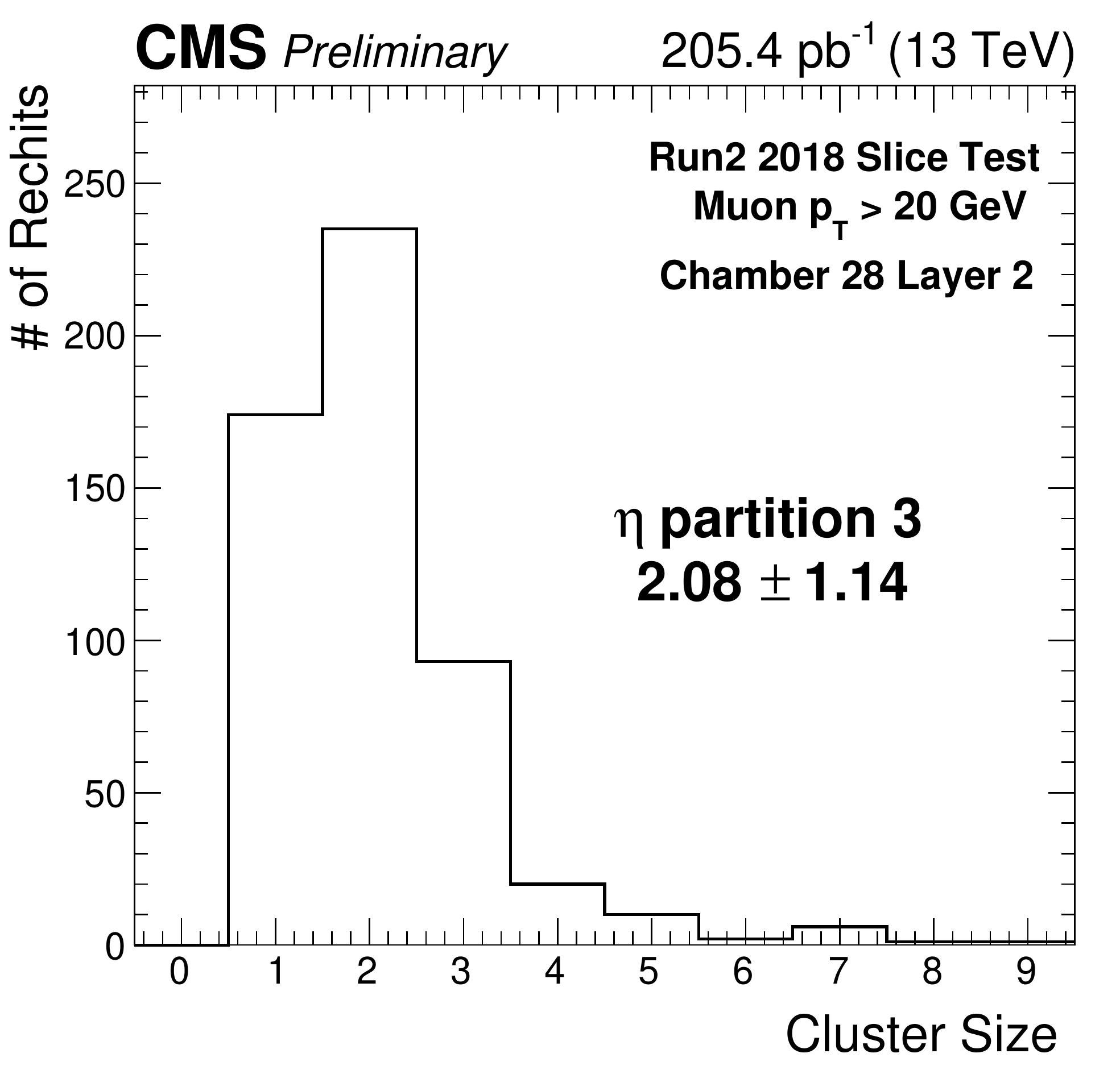}
  \includegraphics[width=.24\textwidth]{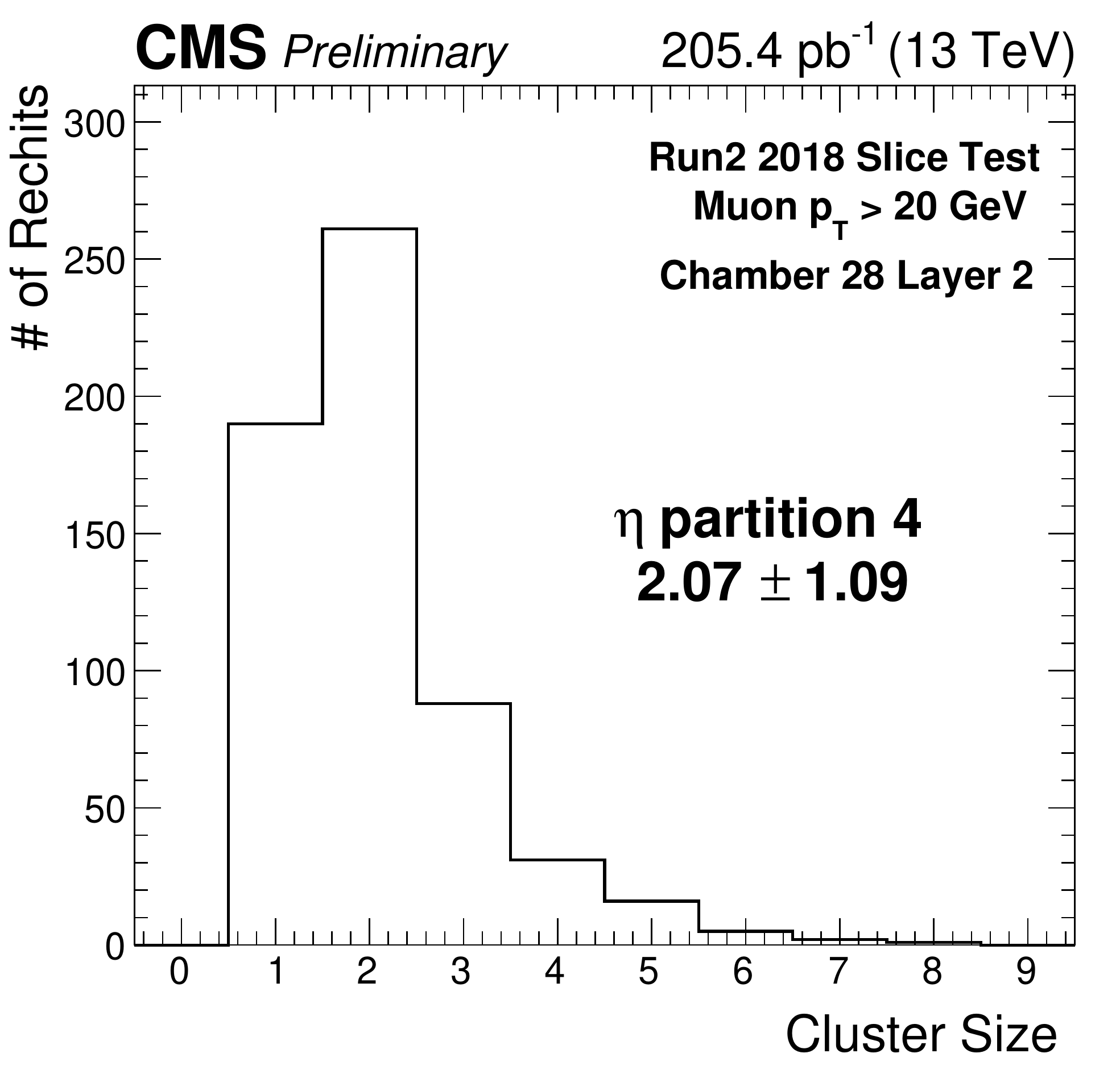}
  
  \includegraphics[width=.24\textwidth]{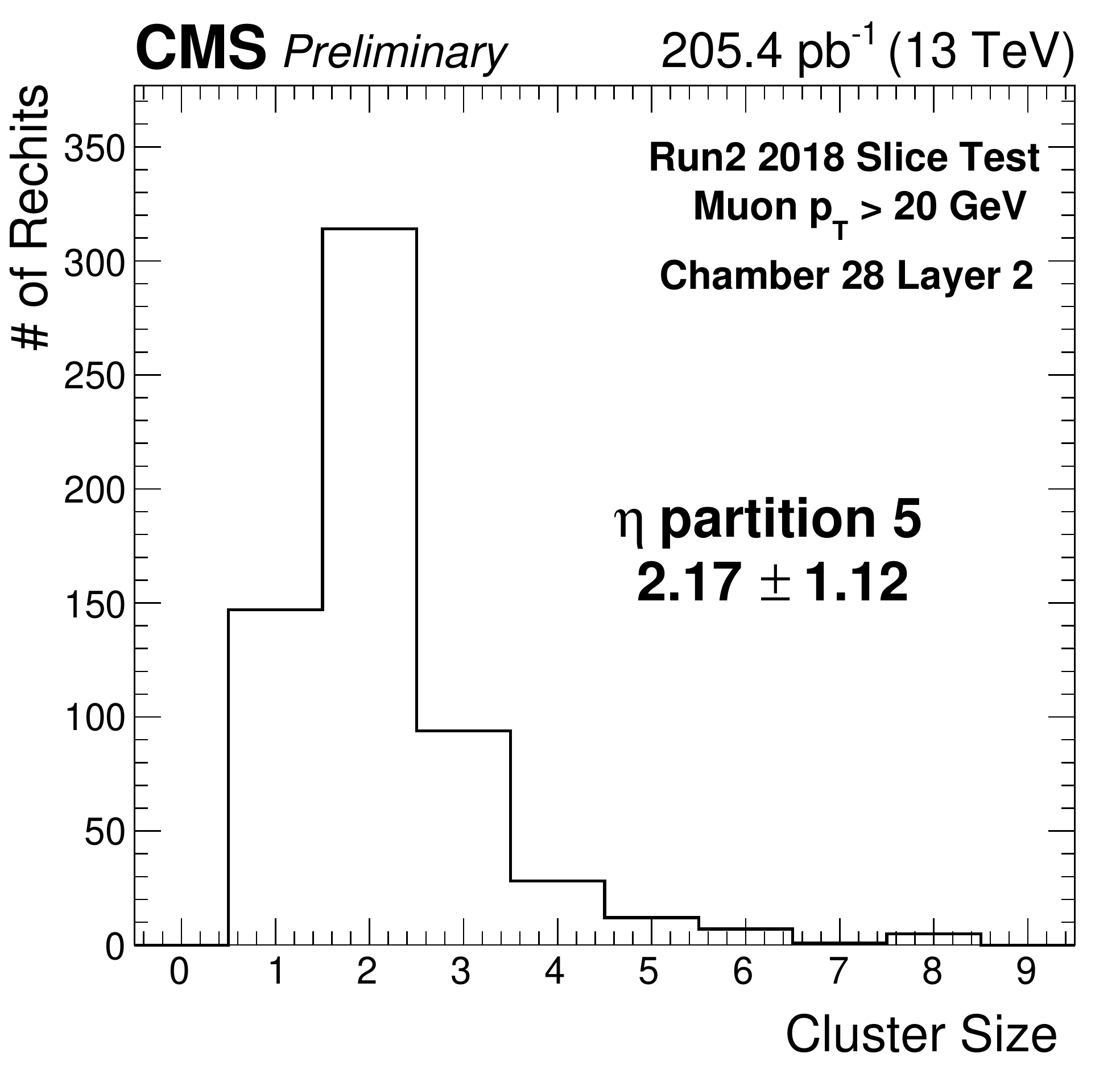}
  \includegraphics[width=.24\textwidth]{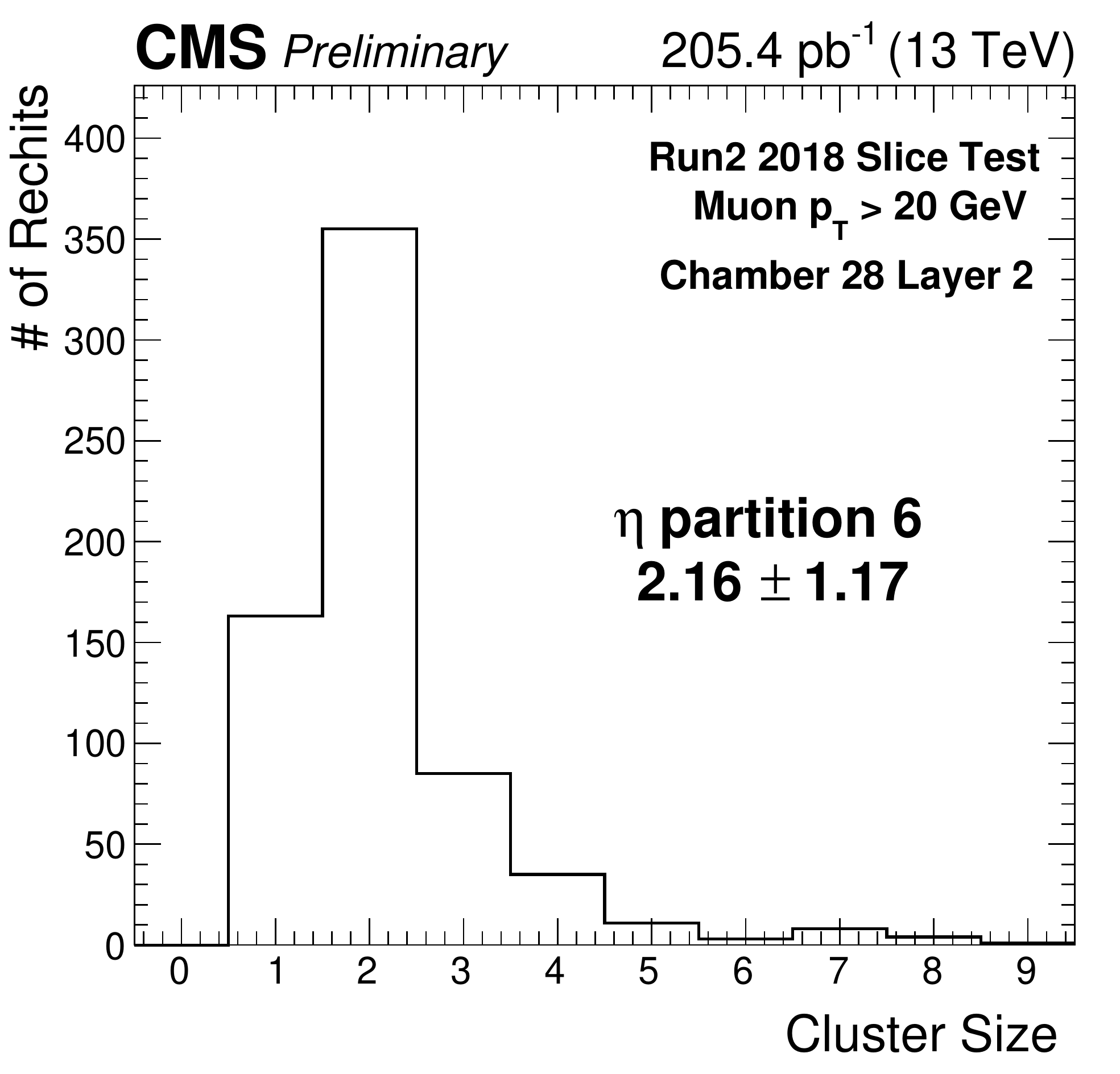}
  \includegraphics[width=.24\textwidth]{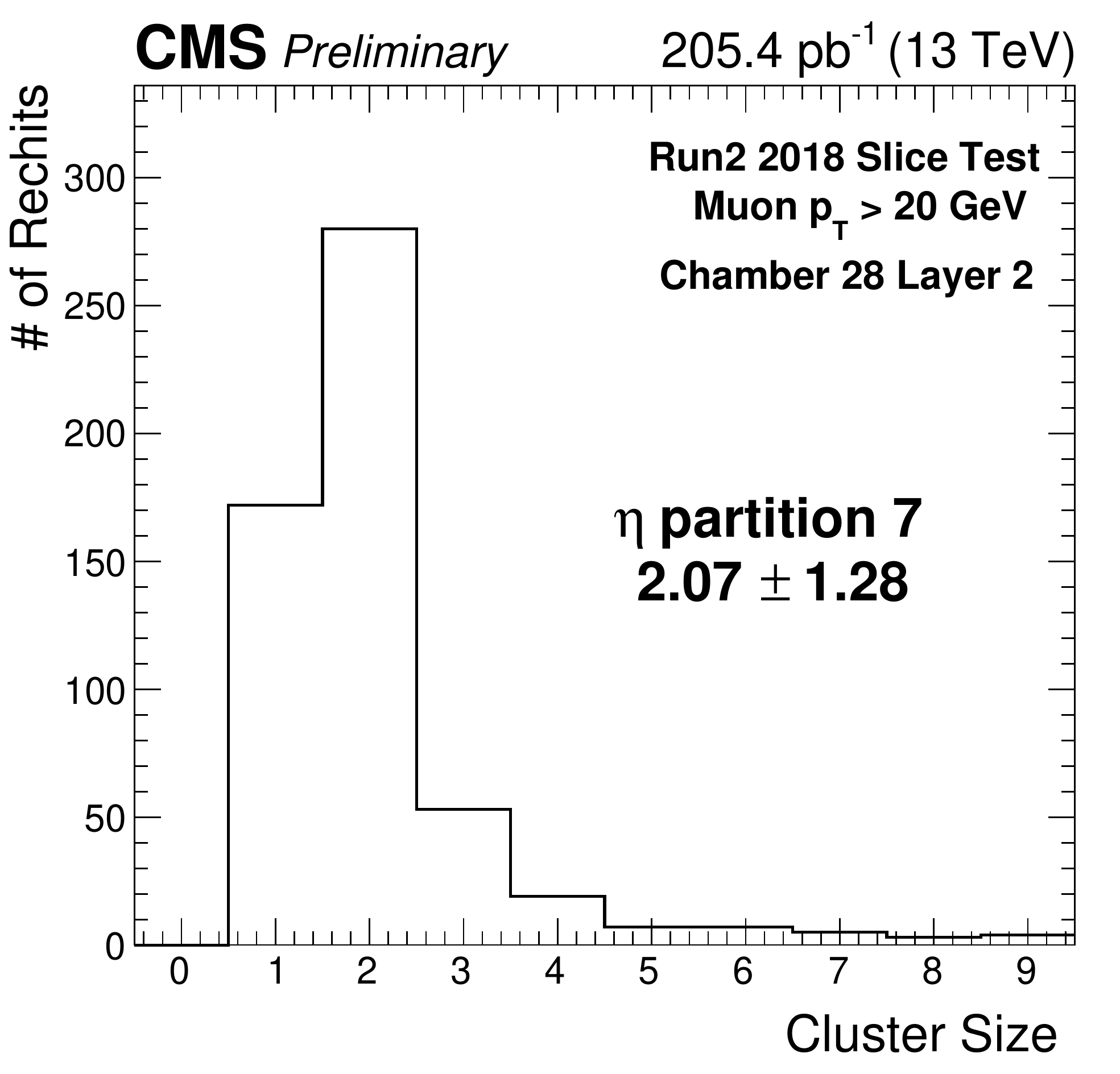}
  \includegraphics[width=.24\textwidth]{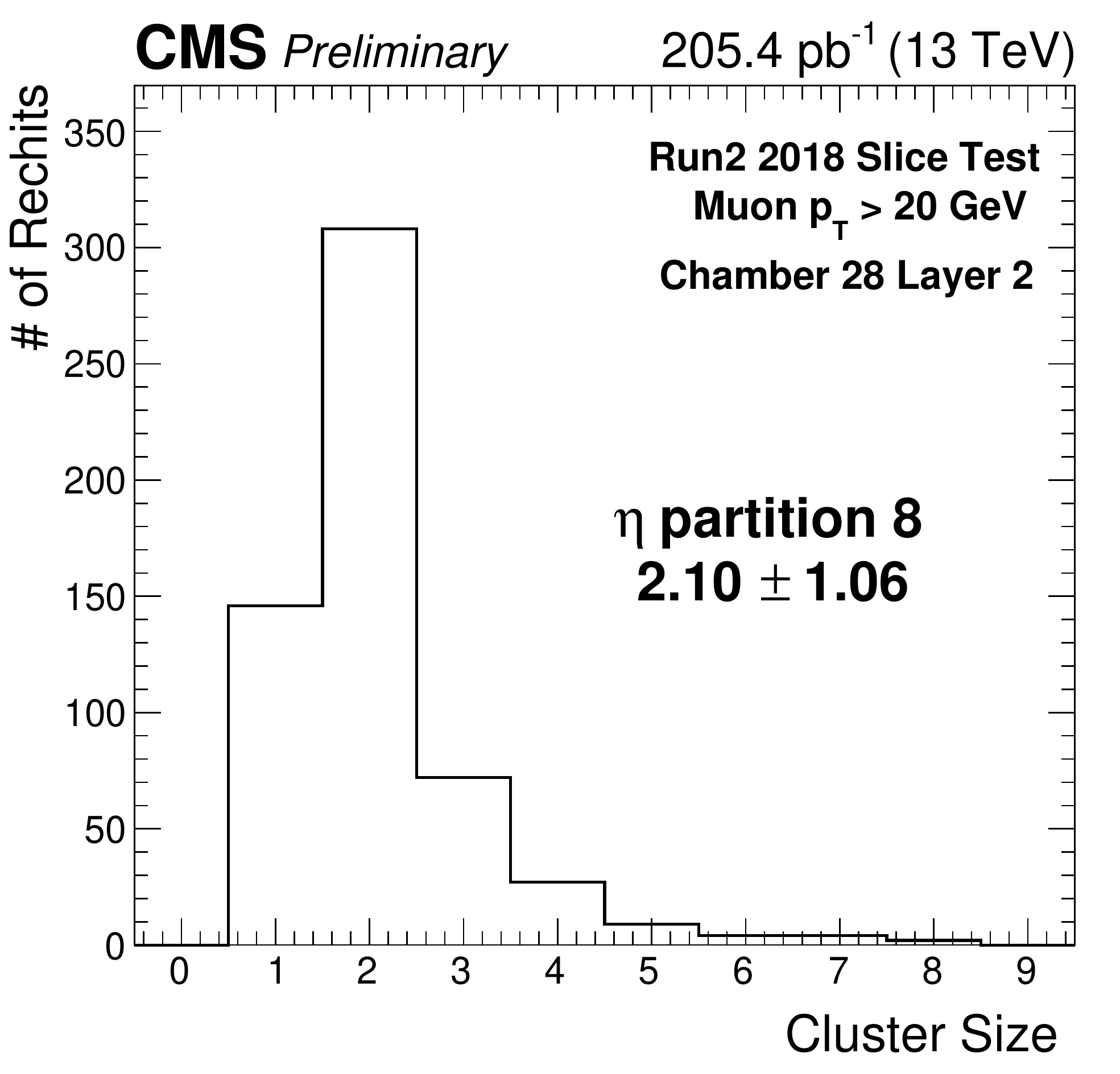}
  \caption{The distribution of cluster size of the GEM RecHits for each of the \(\eta\) partitions in chamber 28, layer 2. 
The \(\eta\) partitions are numbered from 1 at lowest \(\eta\), to 8 at highest \(\eta\). 
The cluster size increases as the \(\eta\) partition number increases as the trapezoidal detector shape means the strips are closer together at high \(\eta\).}
  \label{fig:cls-28-2}
\end{figure}

The data from events passing the CMS triggering system is read out and reconstructed offline using the CMS software, CMSSW.
The raw data from the GEM consists of one bit for each strip which is 1 when a strip fires and 0 otherwise. The data from all strips is packed together with information on the status of the electronics and is unpacked using a map which connects the raw bits to the corresponding strips.
A strip is registered as fired if the charge collected in the timing window (100~ns or four proton--proton bunch crossings in the demonstrator setup) exceeds a charge threshold which is set per $\eta$ partition.
As particles passing through the detector may fire several contiguous strips, groups of adjacent strips which have all fired are clustered into a single reconstructed object, called a GEM RecHit.
This clustering procedure also takes into account strips that are not responsive during the run (inactive strips).
The distribution of cluster size (the number of strips clustered into each RecHit) taken in real data is shown in figure~\ref{fig:cls-28-2} for one of the demonstrator chambers.
Only clusters reconstructed along the path of a muon candidate are included in the distribution, where the muon candidates are chosen with the selection described in section~\ref{sec:muon-det-eff}.
The cluster size is dependent on the gain and the position in the readout, as the wedge geometry of the GEM chambers mean that the strips at high $\eta$ are closer together than at low $\eta$. 
Each detector must pass a requirement of the gain non-uniformity to be less than $\pm15\%$. 
A clear correlation between the muon detector efficiency and its gain non-uniformity is not seen in the demonstrator detectors.
The average of the cluster size was shown to be about two during test beam for all $\eta$ partitions~\cite{Colaleo:2021453}.
This is consistent with the measurements shown here, except for $\eta$ partitions one and two.
These partitions were run with a higher charge threshold due to the grounding issue mentioned earlier, which biased the cluster size to lower values, as seen in figure~\ref{fig:cls-28-2}.
The clusters being readout in conjugation with reconstructed muons and with the expected cluster sizes show the successful operation of the GEM data pipeline from initial data acquisition, to passing through the CMS DAQ and being saved to disk, to unpacking and reading the strip hits, and finally clustering into RecHits.

\section{Muon Detection Efficiency}
\label{sec:muon-det-eff}

The main goal of GE1/1 is to provide triggering and redundancy to the muon tracking system. 
Therefore, understanding the detection efficiency for muons is an important result from the demonstrator detectors.
To understand the in situ response of the GEM detector to muons, we utilize well-identified muons reconstructed with the full CMS detector, except for the GEM detectors.
The muons used in this study are required to pass the tight identification criteria~\cite{collaboration_2012}, and have transverse momentum greater than 20 GeV, from events where the VFATs are reporting the same bunch crossing time as the back end electronics.
We take this collection of muons, then propagate each muon from the final hit to the plane which contains the GE1/1 detectors.
If the muon is successfully propagated inside a GE1/1 detector, then we consider the muon as a GEM-occupying muon.
We additionally require that the muon propagation point is not within $0.5^\circ$ of the edge of the detector, to exclude edge inefficiencies due to uncertainties in the propagation.
We search the $\eta$ partition to which the muon is propagated, and if a GEM RecHit is found within 5 cm of the propagation point, we consider that RecHit matched to the reconstructed muon.

\begin{figure}[t]
  \centering
  \includegraphics[width=.45\textwidth]{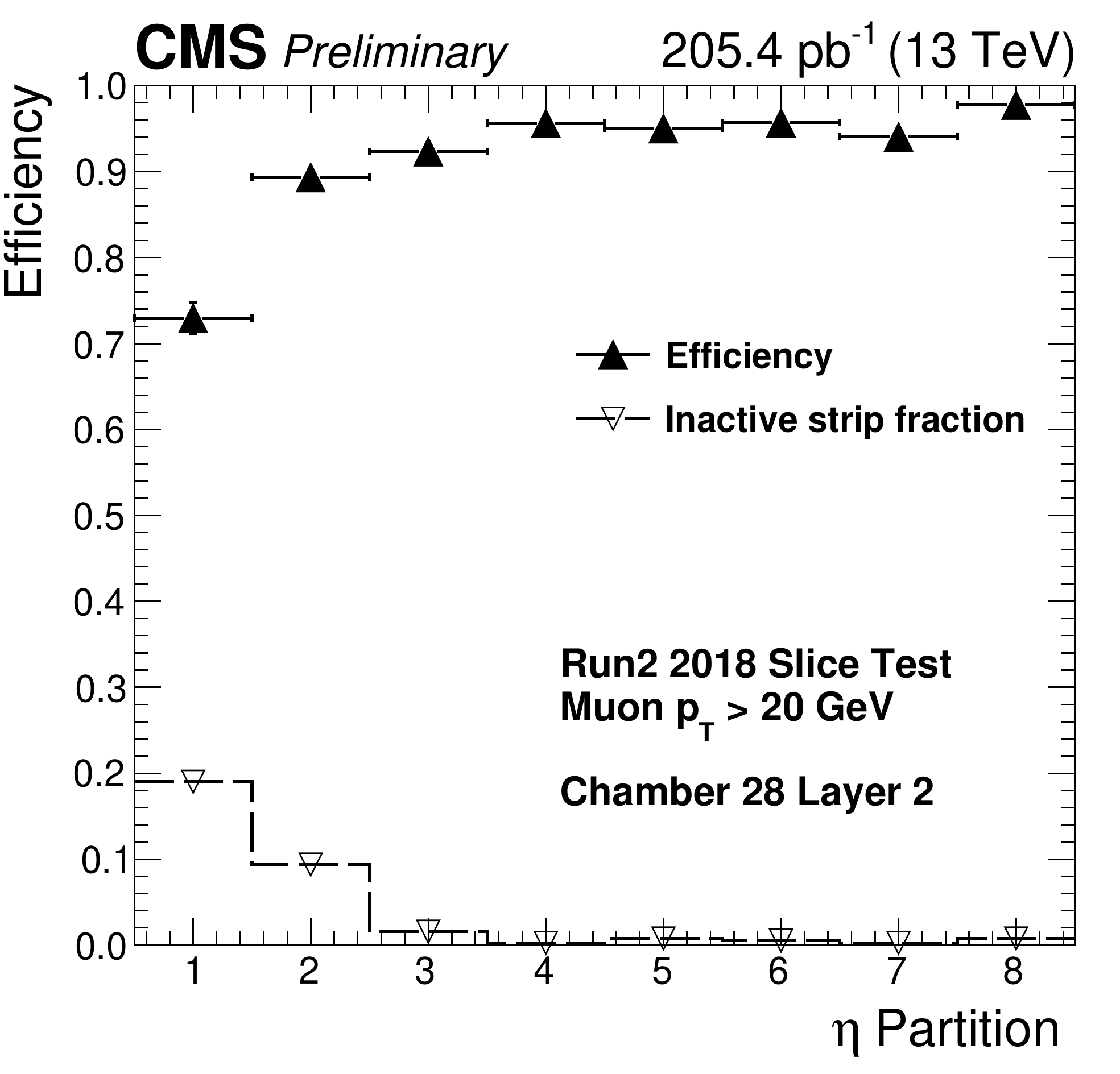}
  \caption{Detector efficiency for the demonstrator chambers as measured using muon reconstructed in each of the \(\eta\) partitions of one of the GEM demonstrator chambers. The inactive strip fraction indicates the fraction of strips unable to collect data during the test period.}
  \label{fig:eff-ch28lay2}
\end{figure}

We take the number of matched GEM RecHit and divide by the total number of muons propagated to the GE1/1 detectors to get the muon detection efficiency.
Figure~\ref{fig:eff-ch28lay2} shows this value for one of the demonstrator chambers, calculated separately for each $\eta$ partition.
Due to electronics issues, some VFATs were inactive and not registering hits during the data-taking period.
The fraction of these inactive channels is indicated and should be taken into account when interpreting the efficiencies, which is taken based on all muons propagated to the $\eta$ partition.
With this caveat in mind, we see that the efficiency of the demonstrator is close to the requirement of 97\% efficiency, defined by the GEM TDR~\cite{Colaleo:2021453}.

\section{Environmental Background Rates}

Another important result from the demonstrator detector is the measurement of the background rate.
That is, the expected rate of particles passing through the GEM detector as a function of luminosity.
This is important as it allows us to understand the level of radiation damage.
We produced measurements of the background rate using a stream of data which was obtained from saving the output of randomly triggered bunch crossings, giving a zero-bias dataset.

We performed several calibration steps to obtain the event hit rate.
To analyze only the active channels, we excluded inactive or noisy channels.
Inactive channels had no hits registered in the entire dataset.
Noisy channels are channels that contain more than 2\% of the total number of hits registered by their VFAT (since there are 128 strips per VFAT we expect each channel to produce about 0.78\% of the VFAT's total hits).
After the removal of the noisy and inactive channels, the remaining channels were used to calculate the active area of the readout by multiplying the total area of the GEM detectors by the fraction of active channels.

Even after removing noisy channels, some luminosity blocks (corresponding to data taking time on the order of a minute) appeared to have all the strips in a single VFAT to fire continuously.
For each VFAT, luminosity blocks where the number of hits is more than \(2\sigma\) above the average are excluded from the event hit rate calculation.
The remaining luminosity blocks define the active time of the VFAT, which is the number of events the VFAT is declared active times the data collection time window of 100 ns (four LHC bunch crossings).
The event hit rate is then calculated as the number of hits divided by the active time and active area.

\begin{figure}[t]
  \includegraphics[width=.55\textwidth]{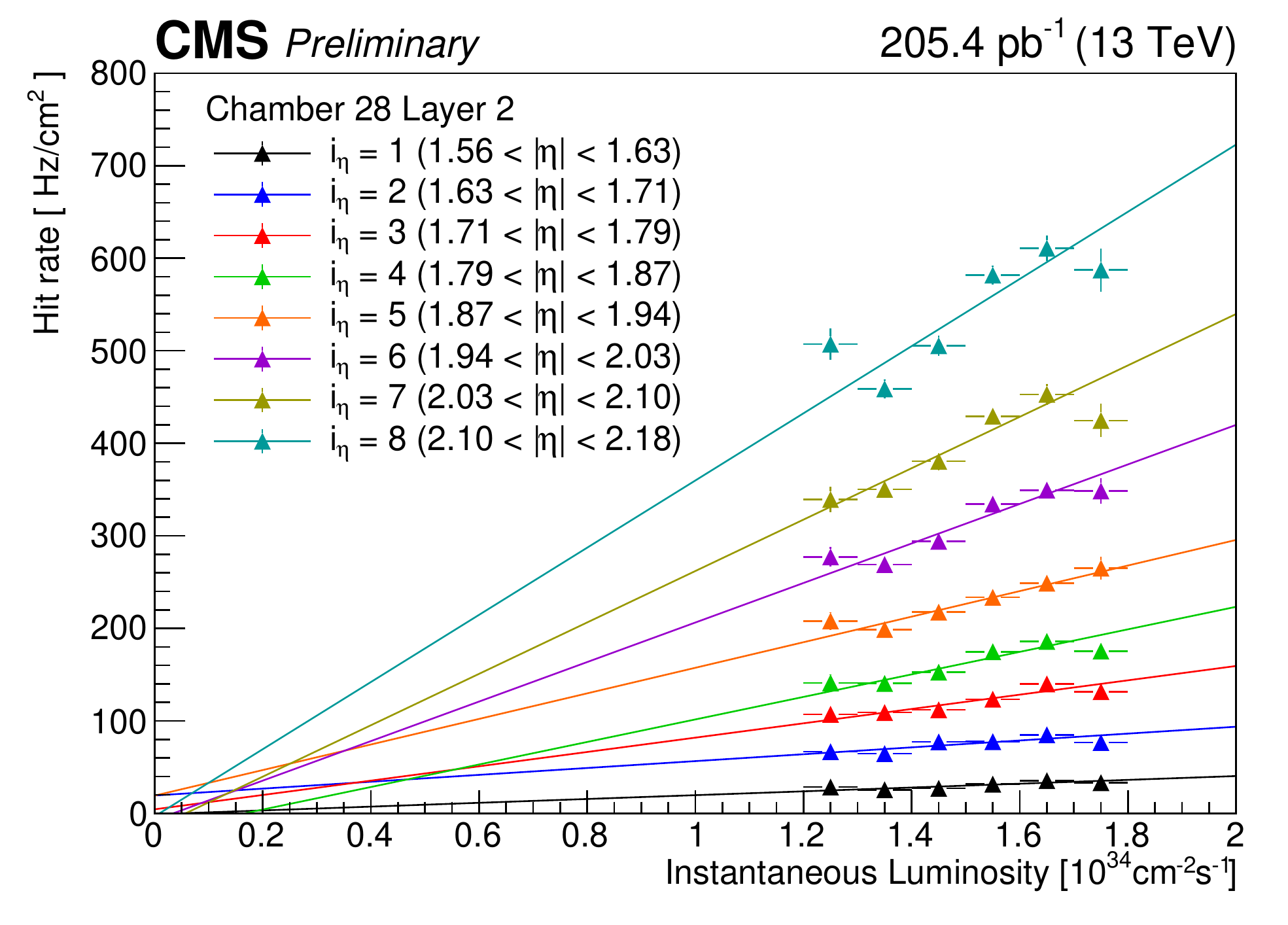}
  \includegraphics[width=.41\textwidth]{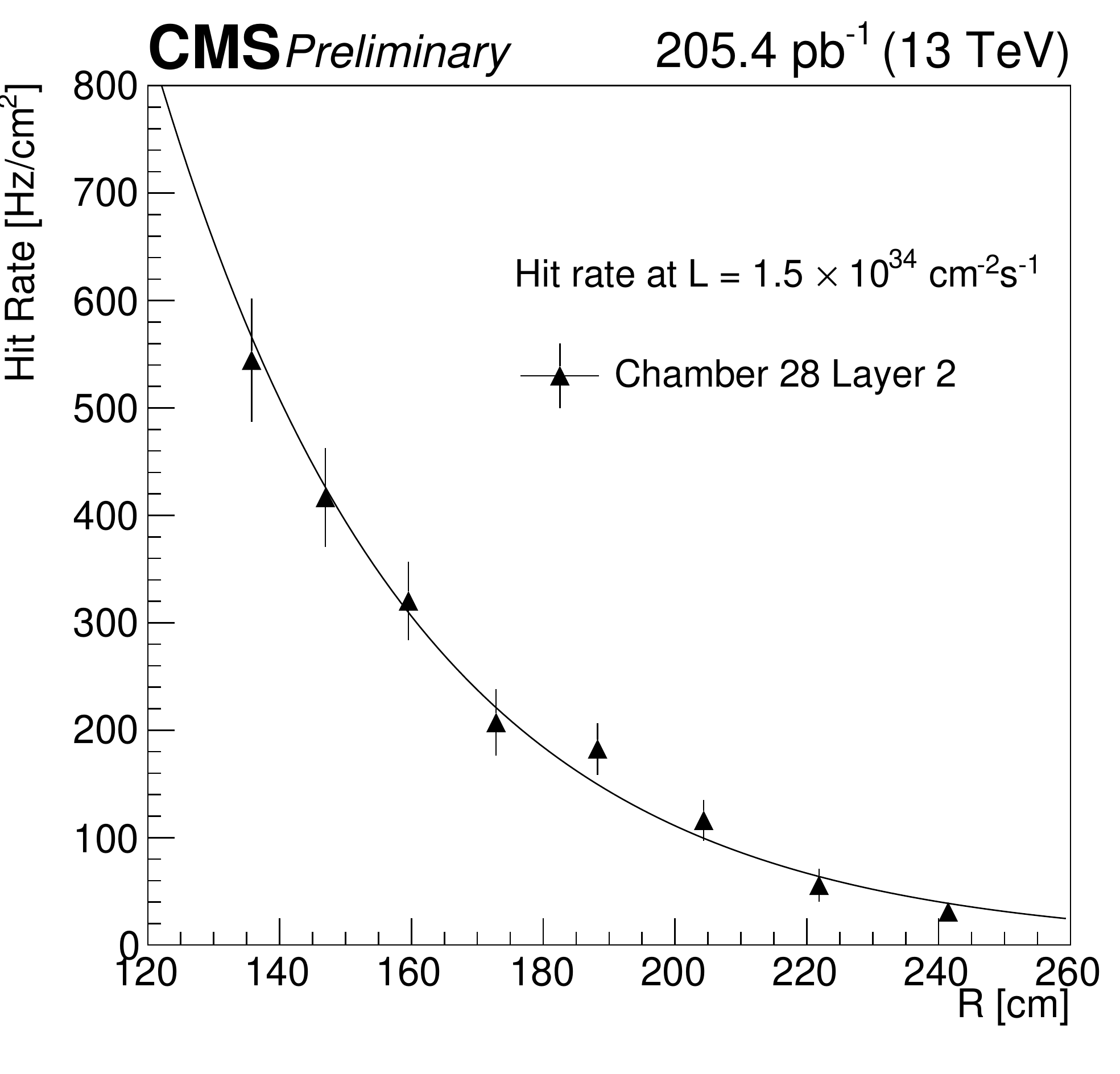}
  \caption{
   Hit rate as a function of the instantaneous luminosity calculated for each $\eta$ partition of chamber 28 layer 2 (left). The data points are corrected for the effective readout area and active time as described in the text.
Hit rate as a function of the distance from the beam pipe where the value of the data points are obtained from the linear fits to hit rate versus instantaneous luminosity of $1.5 \times 10^{34}$ cm$^{-2}$ s$^{-1}$ (right). The error bars are obtained from uncertainties in the linear fit and the curve is an exponential function.
  }
  \label{fig:global-hit-rate}
\end{figure}

Figure~\ref{fig:global-hit-rate} shows the hit rate obtained from the procedure above.
The hit rate is also shown as a function of cylindrical radius for a fixed reference luminosity of $1.5 \times 10^{34}$ cm$^{-2}$ s$^{-1}$, obtained from a linear fit to the binned plot of the rate as a function of luminosity.
In data, the various radii are obtained by using the hit rate for a single \(\eta\) partition and plotting at the average radius, and will, therefore, actually be an average over the radial range of the \(\eta\) partition.
The data show reasonable agreement with the expected background rate based on previous simulation work~\cite{Colaleo:2021453}, and are consistent with newer simulation produced after the data collection period~\cite{simulation-update}.

\section{Summary}

A demonstrator GEM detector was installed into the CMS experiment and operated through 2017 and 2018.
The goal of the detector was to gain operational experience and do performance studies before the full ring of GEM detectors is operational in Run 3 of the LHC.
Through the second half of 2018, the detector was operated as part of the full CMS data acquisition system, allowing for studies to be made using the full CMS data pipeline.
These studies show that the detector has high efficiency for reconstructing hits along moun candidate paths, chosen by the mature CMS identification algorithms.
The rate of background hits was also measured in data.
The measurements performed are consistent with expectations based on previous simulation and test beam studies.
A total of 144 production GE1/1 chambers installed at station 1 of the endcap region will be operational with the expected performance in Run 3 of the LHC.

\begin{acknowledgments}

We gratefully acknowledge support from FRS-FNRS (Belgium), FWO-Flanders (Belgium), BSF-MES (Bulgaria), MOST and NSFC (China), BMBF (Germany), CSIR (India), DAE (India), DST (India), UGC (India), INFN (Italy), NRF (Korea), MoSTR (Sri Lanka), DOE (USA), and NSF (USA). 
\end{acknowledgments}

\bibliography{paper}
\bibliographystyle{JHEP}

\end{document}